# On the Large-Scale Structure of Galactic Disks

**V. P. Reshetnikov***

*Astronomical Institute, St. Petersburg State University, Bibliotechnaya pl. 2, Petrodvorets, 198904 Russia*

**Abstract**—The disk positions for galaxies of various morphological and nuclear-activity types (normal galaxies, QSO, Sy, E/S0, low-surface-brightness galaxies, etc.) on the $\mu_0$–$h$ (central surface brightness–exponential disk scale) plane are considered. The stellar disks are shown to form a single sequence on this plane ($\mathrm{SB}_0 = 10^{-0.4\mu_0} \propto h^{-1}$) over a wide range of surface brightnesses ($\mu_0(I) \approx 12$–25) and sizes ($h \approx 10$–100 kpc). The existence of this observed sequence can probably be explained by a combination of three factors: a disk-stability requirement, a limited total disk luminosity, and observational selection. The model by Mo *et al.* (1998) for disk formation in the CDM hierarchical-clustering scenario is shown to satisfactorily reproduce the salient features of the galaxy disk distribution on the $\mu_0$–$h$ plane. *© 2000 MAIK "Nauka/Interperiodica".*

## 1. INTRODUCTION

In recent years, semianalytical theories of galaxy formation have developed to a point where it has become possible to perform detailed computations of the main parameters of galaxies of various types (e.g., Kauffmann 1996; Dalcanton *et al.* 1997; Mo *et al.* 1998; Jimenez *et al.* 1998; Boissier and Prantzos 1999). Of special interest is the model suggested by Mo *et al.* (1998) (hereafter MMW). This model allows simple analytical formulas to be obtained for the main parameters of galaxy disks and their evolution with $z$ to be estimated.

In this paper, we sum up the observational data on the integrated parameters of the disks of the galaxies of various types (normal galaxies, SyG, QSO, etc.) and compare them with the predictions of the MMW model.

## 2. DISK GALAXIES IN THE MMW MODEL

Mo *et al.* (1998) use the approach that is based on the CDM (Cold Dark Matter) scenario of galaxy formation (see, e.g., the review by Silk and Wyse, 1993). According to this scenario, nonbaryonic halos form from primordial fluctuations at the first stage. At the next stage, gas cools and condenses in these halos to form the disks of the galaxies that surround us.

Here are the main assumptions of the MMW model:

(1) The mass of the galaxy disk ($M_d$) is equal to a fixed fraction ($m_d$) of that of the surrounding dark halo ($M$), i.e., $M_d = m_d M$.

(2) The angular momentum of the disk ($J_d$) is also equal to a fixed fraction ($j_d$) of that of the halo: $(J) - J_d = j_d J$.

(3) Galaxy disks are thin, rotationally supported, and have exponential density distribution $\Sigma(r) = \Sigma_0 \exp(-r/h)$, where $\Sigma_0$ and $h$ are the central disk surface density and exponential disk scale length, respectively.

(4) The disks of real galaxies are stable.

We can show, neglecting self-gravitation of the disk and assuming that the halo is an isothermal sphere, that (see MMW)

$$h = \frac{1}{10\sqrt{2}}\lambda V_c \frac{j_d}{m_d}\left(\frac{H(z)}{H_0}\right)^{-1}, \quad (1)$$

$$\Sigma_0 = \frac{10}{\pi G} m_d \lambda^{-2} V_c \left(\frac{m_d}{j_d}\right)^2 \frac{H(z)}{H_0}, \quad (2)$$

and

$$M_d = \frac{1}{10G} m_d V_c^3 \left(\frac{H(z)}{H_0}\right)^{-1}. \quad (3)$$

In formulas (1)–(3), $\lambda$ is the dimensionless spin parameter defined in the standard way as $\lambda = JE^{1/2}G^{-1}M^{-5/2}$ ($E$ is the total energy; $G$, the gravitational constant); $V_c$, the disk rotational velocity; $H_0$, the current Hubble constant; and $H(z)$, the Hubble constant at redshift $z$ corresponding to the formation epoch of the dark matter of the halo where the disk was born. Naturally, $H(z)$ depends on the adopted cosmological model. Hereafter, all computations are made in terms of the flat cosmological model ($\Omega = 1$ and $\Omega_\Lambda = 0$) with $H_0 = 75$ km s$^{-1}$ Mpc$^{-1}$. The result of the allowance for disk self-gravitation and the use of more realistic dark halo mass distributions is a factor on the order of unity in the above formulas (MMW). Hereafter, we neglect this factor.

* E-mail address for contacts: resh@astro.spbu.ru

It follows from formulas (1) and (2) that the total disk luminosity is $L_d \propto \Sigma_0 h^2 \propto V_c^3$. The MMW model thus has a Tully–Fisher relation already incorporated into it [see a discussion in Mo *et al.* (1998) and Boissier and Prantzos (1999)].

According to Mo *et al.* (1998), the main properties of disk galaxies are fully determined by parameters $\lambda$, $m_d$, $j_d$, $V_c$, and $H(z)$. $H(z)$ increases with $z$, and, therefore, as is evident from formulas (1) and (2), the disks that formed at large $z$ should be denser and more compact, other parameters being fixed. Higher $\lambda$ must correspond to more extended and relatively more rarefied disks.

Other authors also have given analytical formulas for the parameters of disk galaxies. Thus, for example, Dalcanton *et al.* (1997) found that $h \propto \lambda M^{1/3}$ and $\Sigma_0 \propto \lambda^{-2(1+3F)} M^{1/3}$, where $F$ is the baryonic mass fraction of the galaxy. For small $F$, the model of Dalcanton *et al.* (1997) implies a dependence of disk parameters on $\lambda$ and $M$ that is close to that implied by MMW. Similar relations were found even earlier by van der Kruit (1987). However, Mo *et al.* (1998) wrote these relations in a form that is more convenient to use in the analyses. Furthermore, the relations given by the latter authors include explicit dependence of disk parameters on the disk formation time.

We now analyze how the data on real galaxies of various types agree qualitatively and quantitatively with predictions of the model of Mo *et al.* (1998).

## 3. GALAXIES OF VARIOUS TYPES ON THE $\mu_0$–$h$ PLANE

### 3.1. Normal Galaxies

In his dissertation, Byun (1992) gives the results of a two-dimensional decomposition of the surface-brightness distributions for 1163 Southern-sky Sb–Sd spiral galaxies in the Kron–Cousins $I$ band. Small circles in Fig. 1 show the distribution of the disks of these galaxies on the $\mu_0(I)$–$h$ plane (here $\mu_0(I)$ is the central disk surface brightness of the disk in magnitudes per square arcsecond; and $h$, the exponential scale length in kpc assuming $H_0$ = 75 km s$^{-1}$ Mpc$^{-1}$). Filled circles indicate the galaxy disks located within 15 Mpc from us. The latter subsample contains a larger fraction of galaxies with relatively small $h$, which can be naturally explained by observational selection (the sample of Byun (1992) is angular diameter-limited, because it contains only the galaxies with $d \geq 1'.7$).

The mean parameters of the galaxy sample considered are $\langle\mu_0(I)\rangle = 19.43 \pm 0.83(\sigma)$ (uncorrected for internal extinction within sample objects) and $\langle h \rangle = 3.9 \pm 1.8(\sigma)$ kpc. Adopting $B - I = +1.70$ for galaxy disks (de Jong 1996b), we obtain $\langle\mu_0(B)\rangle \approx 21.1$, which is somewhat brighter than the standard value of 21.65 (Freeman 1970) due to observational selection. Byun (1992) showed the galaxy distances in his sample to correlate with central disk brightness, i.e., the galaxies with higher $\mu_0$ were selected predominantly among more distant objects. The mean $\mu_0(I)$ for the objects within 30 Mpc is $19.60 \pm 1.00$ (261 galaxies) and thus becomes closer to the standard Freedman's value.

Note also that the parameters of our own Galaxy, the Milky Way, are close to the corresponding mean values for Byun's (1992) sample of spiral galaxies. According to Sackett (1997), $h^{MW} = 3 \pm 1$ kpc. This result, combined with modern estimates of the surface brightness of the Milky Way in the solar vicinity (van der Kruit 1986; Kimeswenger *et al.* 1993), yields $\mu_0^{MW}(B) \approx 21$.

Evstigneeva and Reshetnikov (1999) suggested that the observed parameters of the disks of spiral galaxies obey the relation $SB_0 \propto h^{-1}$, where $SB_0 = 10^{-0.4\mu_0}$. The disk parameters from Byun (1992) do not show such a correlation. The linear correlation coefficient between $\mu_0(I)$ and $\log h$ for the entire sample (1163 galaxies) is equal to 0.06. However, if we restrict the sample to the objects located within 30 Mpc, the correlation coefficient increases to $\rho = 0.20$. Note also that the range of galaxy disk parameters in the sample by Byun (1992) is not too large (see Fig. 1). The dashed line in Fig. 1 shows the $SB_0 \propto h^{-1}$ relation for the data points corresponding to the mean parameters of the galaxies from Byun (1992). As is evident from Fig. 1, extremely compact nuclear stellar disks of elliptical galaxies, the disks of E/S0 galaxies, and very extended disks of giant low-surface brightness galaxies lie more or less along the relation mentioned above (see below).

The observed distribution of normal-galaxy disks on the $\mu_0$–$h$ plane can be explained in terms of simple hypotheses of observational selection and physical disk stability (see e.g., McGaugh 1998; van den Bosch 1998). The galaxy luminosity function falls off abruptly at high absolute luminosities: very luminous galaxies are extremely rare. Therefore, the upper limit of the total galaxy disk luminosity is the most natural constraint in Fig. 1. The thick solid line in Fig. 1 shows the locus of constant exponential disk luminosity provided that the total luminosity is $L_I = 10 L_I^*$. Here, $L_I^*$ corresponds to the characteristic absolute magnitude of $M_I^* = -21.8$ in the galaxy luminosity function according to Marzke *et al.* (1998). It is evident from Fig. 1 that this straight line sets a reasonable upper limit for galaxy disks from Byun (1992).

The second natural constraint for galaxy disks in Fig. 1 is the angular diameter limitation. For exponential disks, we have $\mu(r) = \mu_0 + 1.086r/h$ and, consequently, the limitation on diameters can be rewritten as $d \geq d_{lim} = 2r_{lim} = 1.842h\,(\mu_{lim} - \mu_0)$. The dotted curve in Fig. 1 shows the observational selection relation for a galaxy with $d_{lim}$ = 5 kpc and $\mu_{lim}(I)$ = 23.3 (this limiting $I$-band isophote corresponds to $\mu_{lim}(B) = 25.0$ for $B - I = +1.70$). As is evident from Fig. 1, the galaxies from Byun (1992) are located in the $L_d \le 10L^*$ and $d \ge 5$ kpc domain on the $\mu_0$–$h$ plane. [An apparent magnitude selection produces a relation that is similar to that resulting from angular diameter selection (McGaugh, 1998).]

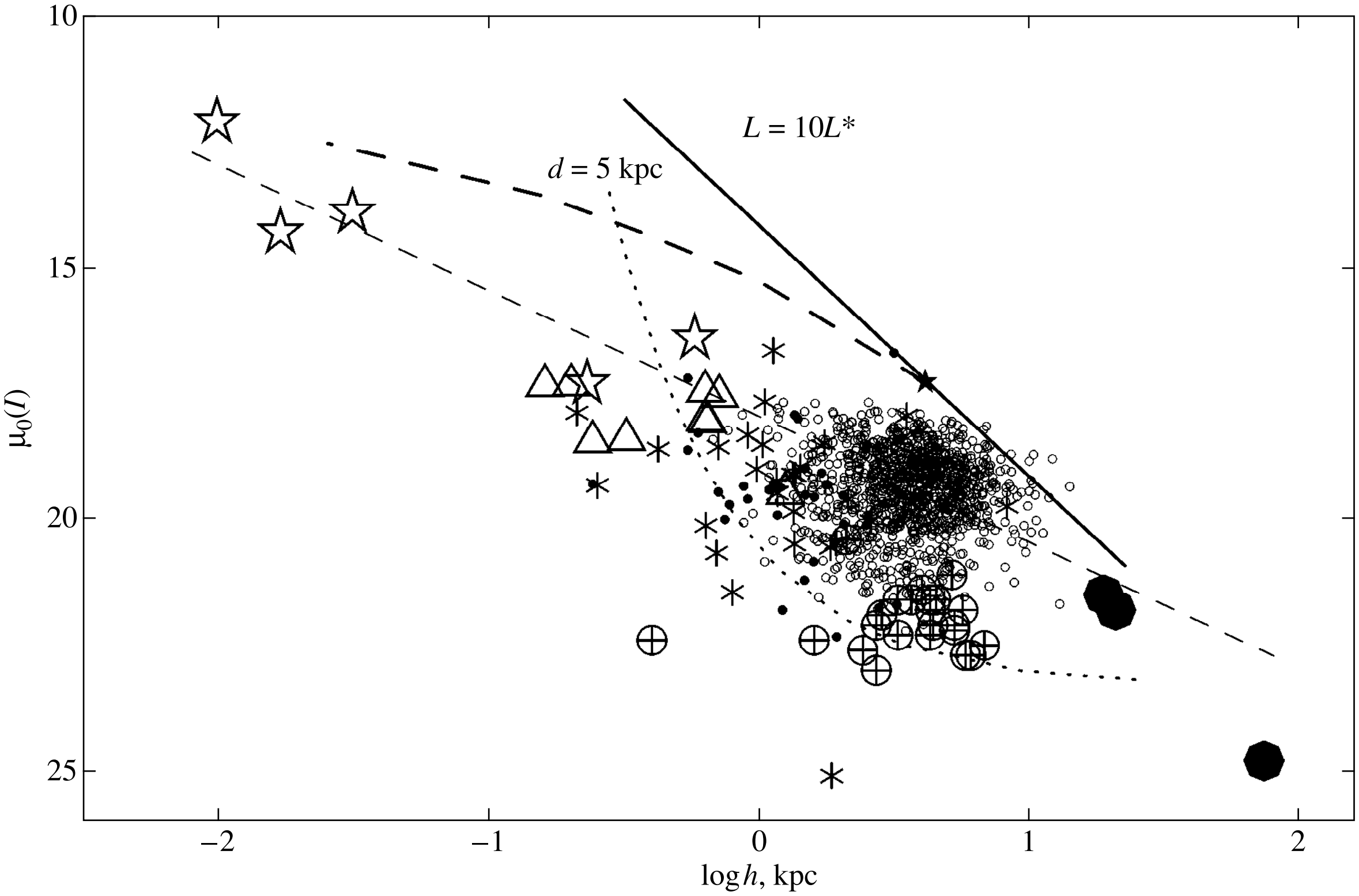

**Fig. 1.** The distribution of the exponential disks of galaxies of various types on the $\mu_0(I)$–$h$ plane. Small circles show the data from Byun's (1992) thesis (filled circles are the galaxies with distances <15 Mpc); stars are compact nuclear disks of E/S0 galaxies; triangles and small asterisks are stellar disks in E/S0 galaxies; circles with crosses, low surface brightness galaxies; and big filled circles, giant low surface brightness galaxies (see text). The thick solid line shows the constant disk luminosity curve (provided that $L_d = 10L^*$); dotted line, the selection line for galaxies with a diameter of 5 kpc; dashed curve, the disk stability condition for the galaxies with a total luminosity of $10L^*$; and dashed line, the $SB_0 \propto h^{-1}$ relation (see text).

Another constraint for galaxy parameters in Fig. 1 follows from the bar instability of the disks of galaxies with spins below certain critical value $\lambda_{crit}$. (See a discussion in Mo *et al.* (1998) and van den Bosch (1998).) Following the approach suggested by van den Bosch (1998), we show on the constant-luminosity line in Fig. 1 the parameters of a disk with $L_d = 10L^*$ and $\lambda = \lambda_{crit} = 0.05$ (MMW) (asterisk). To be stable, more compact disks require a bulge. The thick dashed line in Fig. 1 is the locus where the conditions $L_d + L_b = 10L^*$ (here $L_b$ is the luminosity of the galaxy bulge) and $\lambda = \lambda_{crit}$ are satisfied. The bulge-to-disk luminosity ratio varies along this curve from 0 (asterisk) to $+\infty$. (Note that the exact position of this dashed line depends on the adopted cosmological model and the galaxy formation redshift, $z_f$. We adopt $z_f = 0$ in our illustrative calculations).

It can be easily seen from Fig. 1 that the stability condition combined with luminosity constraint ($L_d \le 10L^*$) determines a domain with an upper limit given approximately by straight line $SB_0 \propto h^{-1}$ on the $\mu_0$–$h$ plane.

### 3.2. *Galaxies with Active Nuclei and Starburst Galaxies*

The thick cross in Fig. 2 marked by letter Q shows the mean disk parameters for 10 relatively nearby quasars (with redshifts $z \approx 0.2$) as inferred from the data published by Bahcall *et al.* (1997)[1]. (We considered only the quasars whose two-dimensional surface brightness distributions can be better fitted by the exponential rather than the deVaucouleurs law.) The central disk surface brightness of the QSO sample objects is somewhat higher (by $\approx 0\overset{m}{.}6$–$0\overset{m}{.}7$) than that of normal galaxies from Byun (1992). However, as is evident from Fig. 2, this discrepancy is within the scatter of disk parameters for normal galaxies.

Crosses S and SS in Fig. 2 show the mean parameters for the Seyfert galaxy sample according to Afanas'ev *et al.* (1997) and for the sample of galaxies with Seyfert and starburst nuclei according to Hunt

[1] Hereafter we assume $B - I = +1.7$ and $V - I = +1.0$ in all transformations between photometric bands

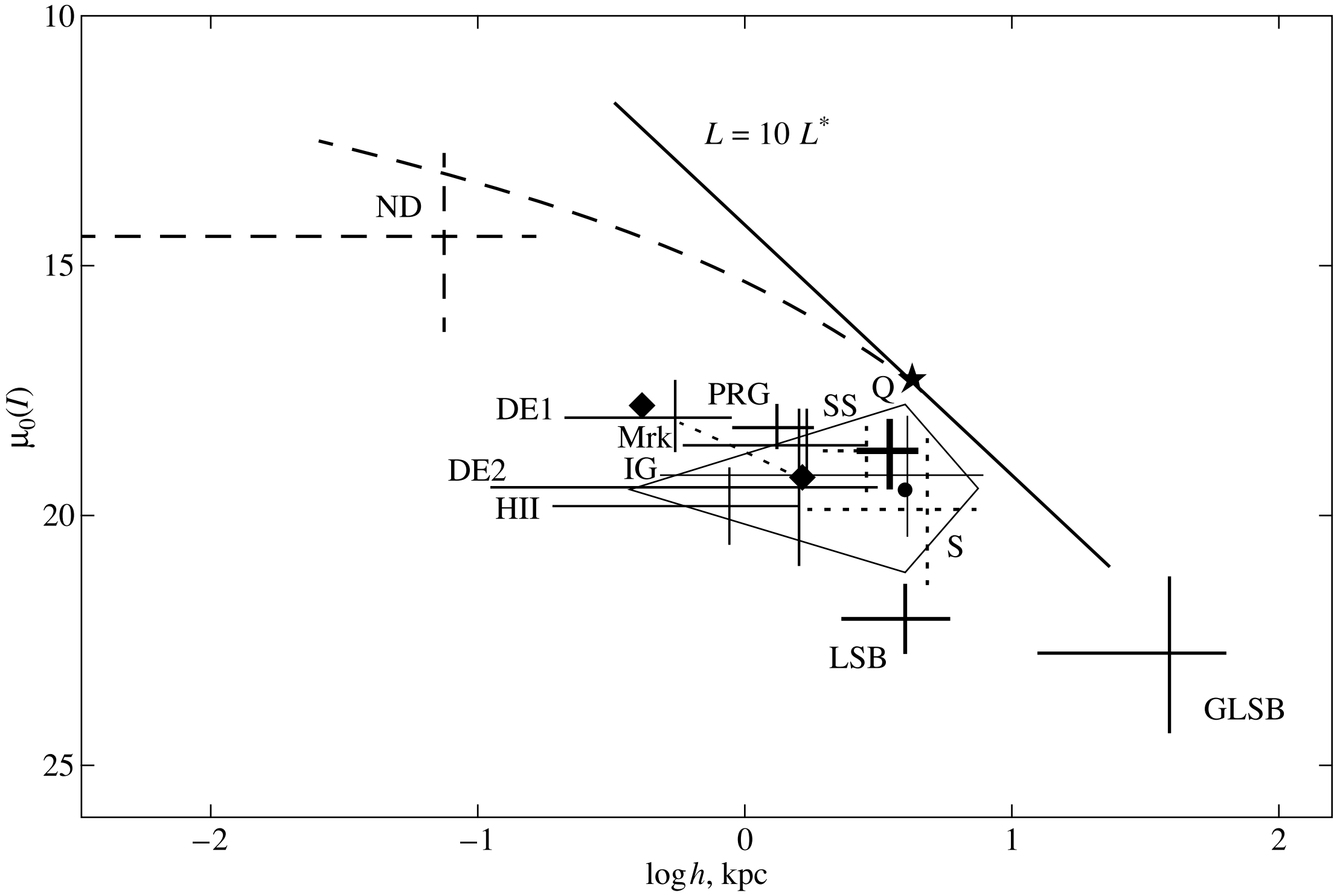

**Fig. 2.** Stellar disks of galaxies of various types on the $\mu_0(I)$–$h$ plane (see text). The dark dot indicates the mean parameters of spiral galaxies from Byun's (1992) thesis, and the quadrangle of the segments of solid lines shows the $\pm 2\sigma$ box of the mean values. The curves have the same meaning as in Fig. 1.

*et al.* (1999), respectively. (We transformed the data taken from Hunt *et al.* (1999) into the $I$ band assuming $I - K = +1.8$.) The mean disk parameters for both samples resemble those of normal galaxies, although, as was pointed out by Hunt *et al.* (1999), the disks of Seyfert-I galaxies are relatively compact and have abnormally high $\mu_0$.

The starburst galaxies from Chitre and Joshi (1999) (all 10 objects are Markarian galaxies; therefore, their parameters are marked by Mrk in Fig. 2) have more compact and brighter disks compared to the common Sb–Sd galaxies. However, the objects from Chitre and Joshi (1999) are mainly early-type galaxies whose disk parameters are known to be shifted in just this direction (see, e.g., de Jong 1996a).

Dwarf starburst galaxies (HII in Fig. 2) from Telles and Terlevich (1997) have relatively low central surface brightness compared to other galaxies of similar size. This corroborates the conclusion by Telles and Terlevich (1997) that HII galaxies are similar to low-surface-brightness galaxies (see Section 3.6).

### *3.3. Interacting Galaxies*

The parameters of interacting galaxies from Evstigneeva and Reshetnikov (1999) (IG in Fig. 2) resemble very much those of normal galaxies as inferred by Byun (1992). However, the mean $B$-band $\mu_0$ for the disks of interacting galaxies is about $1^m$ brighter than that of isolated objects (Evstigneeva and Reshetnikov, 1999). The absence of such a large difference between the $\mu_0$ values in the $I$ band is consistent with the assumption that high surface brightness of interacting disks is due to enhanced star formation. On the other hand, as was shown in Section 3.1, the sample of Byun (1992) is distorted by observational selection, which somewhat shifts the mean $\mu_0$ toward higher values. Observational selection taken into account, the central disk surface brightness of normal galaxies differs from that of interacting galaxies by $\Delta\mu_0(I) \approx 0\overset{m}{.}5$.

Polar-ring galaxies (PRG) from Reshetnikov *et al.* (1994) have relatively compact and bright stellar disks. This can be due both to the real mass redistribution in strong interaction, accompanied by the formation of a polar structure, and to observational selection. (A polar ring lives much longer around a short disk than around an extended one.)

In Fig. 2, diamond signs connected by a dashed line show the parameters of two stellar disks of the Sa-type galaxy NGC 3593 (Bertola *et al.* 1996). The smaller of these disks rotates in the direction opposite that of the larger disks and of the galaxy as a whole. The counter-rotating disk must have formed as a result of slow gas accretion from the intergalactic space onto the gas-free spiral galaxy (Bertola *et al.* 1996). The subsequent star formation in the captured gas must have resulted in the formation of a counterrotating stellar disk. Interestingly, the parameters of this "secondary" disk are close to those of the disks in E/S0 galaxies from Scorza and Bender (1995).

### *3.4. Elliptical Galaxies and Galaxies with Stellar Disks*

In Figs. 1 and 2, stellar disks of nine E/S0 galaxies from Scorza and Bender (1995) (DE1, triangles) and 28 E/S0 galaxies from Scorza *et al.* (1998) (asterisk) lie approximately along the $SB_0 \propto h^{-1}$ relation slightly shifted toward lower surface brightness values (or lower *h*).

Large asterisks in Fig. 1 and the dashed cross (ND) in Fig. 2 show the parameters of compact stellar disks in the nuclei of E/S0 galaxies according to Scorza and Bender (1990), Kormendy *et al.* (1996), and Scorza and van den Bosch (1998). As is evident from Figs. 1 and 2, compact nuclear disks lie along the same continuous sequence on the $\mu_0$–$h$ plane as stellar disks of early-type galaxies and those of spiral galaxies. The location of these disks on the plane considered agrees with the constraint implied by stability considerations—see Section 3.1 (dotted line). It can be assumed that early-type galaxies may contain disks with high $\mu_0$ (they must fill the lower left corner in Figs. 1 and 2); however, they are difficult to identify against the bright E/S0 galaxy background.

### *3.5. Low Surface Brightness Galaxies*

Low-surface-brightness galaxies from de Blok and McGaugh (1997) (crossed circles and LSB-marked cross in Figs. 1 and 2, respectively) have substantially lower surface brightnesses and the sizes that are comparable to those of normal galaxies. Giant low surface brightness galaxies (large filled circles in Fig. 1 and GLSB in Fig. 2) have colossal sizes with exponential disk scale lengths on the order of several tens of kpc (Bothun *et al.* 1987, 1990; Sprayberry *et al.* 1993). "Extreme" stellar disks (ultracompact nuclear stellar disks and GSLB galaxies) approximately follow the $SB_0 \propto h^{-1}$ relation.

Low surface brightness galaxies usually cannot be found in common galaxy catalogs. Such galaxies are usually searched for using special observing equipment, and many such objects still remain undiscovered even in the part of the nearby part of the Universe (see, e.g., Sprayberry *et al.* 1997). Future surveys are expected to yield new LSB galaxies to fill the lower right corners in Figs. 1 and 2.

### *3.6. Main Observational Conclusions*

The disks of real galaxies of various types span a very large range of surface brightnesses ($\mu_0(I) \approx 12$–$25$) and sizes ($h \approx 10$ pc–100 kpc) in the $\mu_0$–$h$ plane, where they form a continuous sequence. The bulge to disk luminosity ratio ($B/D$) varies systematically along this sequence from $\gg 10$ for ellipticals located at the top left corner in Figs. 1 and 2 to ~0 for spiral galaxies located at the bottom right corner in the same figures. The effective bulge radius, $r_e$, to the disk scale length, $h$, ratio varies from $r_e/h \sim 0.1$ for giant low surface brightness galaxies at the bottom right corner in Figs. 1 and 2 to $r_e/h \geq 5$–$10$ for stellar disks in E/S0 galaxies. $\mu_0$ and $h$ also vary along the same sequence: the disks of galaxies with higher $B/D$ ratios are, on the average, brighter and more compact.

The disks of quasars, Seyfert galaxies, and galaxies in interacting systems might have somewhat higher-than-average $\mu_0$ values, due possibly to an increased star-formation rate and real mass redistribution in such galaxies. Note, however, that one should be very careful when concluding that the parameters of the disks of the galaxies of a certain type differ from those of "normal" galaxies because of selection effects that are difficult to assess. (The samples compared must satisfy similar apparent magnitude and angular size limitations and have similar B/D.) Furthermore, the structure of the disks can differ systematically in the regions with different space density of galaxies (Moore *et al.* 1998). On the whole, as is evident from Figs. 2 and 3, the parameters of the disks of galaxies of various morphological and activity types agree with those of the disks of normal galaxies, which are here represented by the objects from Byun's (1992) thesis. The compact stellar disks in the nuclei of E/S0 galaxies and low surface brightness galaxies have parameters that differ most from those of normal galaxies.

The observed positions of galaxies on the $\mu_0$–$h$ plane are determined by the disk stability condition, the total disk luminosity constraint ($L_d \leq 10L^*$), and observational selection. A combination of these three conditions (and, perhaps, other unknown factors) results in the alignment of galaxy-disk parameters along the empirical $SB_0 \propto h^{-1}$ relation (see Figs. 1–3). Deeper (in terms of surface brightness) stellar disk surveys are likely to "blur" this observational relation (see, e.g., Fig. 4 in Dalcanton *et al.* 1997), which, however, should survive in the form of the upper envelope of galaxy parameter distribution.

## 4. THE MO *et al.* (1998) MODEL AND REAL GALAXIES

### *4.1. The Milky Way*

The integrated properties of disk galaxies in the Mo *et al.* (1998) model depend on a number of parameters (see Section 2)—$\lambda$, $m_d$, $j_d$, $V_c$, and redshift, $z_f$,—of the dark halo (disk) formation. We now estimate the time scale of the Milky Way disk formation in terms of the model of Mo *et al.* (1998) using realistic parameter values. We adopt $\lambda = 0.05$ and $m_d = j_d = 0.05$ (MMW) as standard values. Then, assuming for the Milky Way disk $h^{MW} = 3$ kpc; $\mu_0^{MW}(B) \approx 21$, $V_c^{MW} = 200$ km s$^{-1}$, and $M_d^{MW} = 6 \times 10^{10} M_\odot$ (see Section 3.1; Reshetnikov 2000), we obtain from formulas (1)–(3) that $H(z)/H_0 \approx 3$ (adopting a $M_d/L_d(I) = 1.7$ mass-luminosity ratio for the disk). It follows from this that $z_f \approx 1$ for $\Omega = 1$ and $\Omega_\Lambda = 0$ (see Fig. 1 in Mo *et al.* 1998). The model of Mo *et al.* (1998) therefore yields self-consistent parameter estimates for the Milky Way disk and its formation epoch ($z_f \approx 1$). Note also that the $m_d = 0.05$ value adopted above yields a plausible dark halo mass estimate for the Milky Way, $M = M_d/m_d = 1.2 \times 10^{12} M_\odot$ (see, e.g., the review by Zaritsky 1998). Note, however, that the dynamical analysis of the Galactic companions, the Galaxy + M31 system, etc, imply a larger size (~200 kpc) and dark-halo mass ($>10^{12} M_\odot$) for our Galaxy. Local observations provide no evidence for considerable dark mass within the optical disk of the Milky Way (e.g., Kuijken and Gilmore 1991; Creze *et al.* 1998).

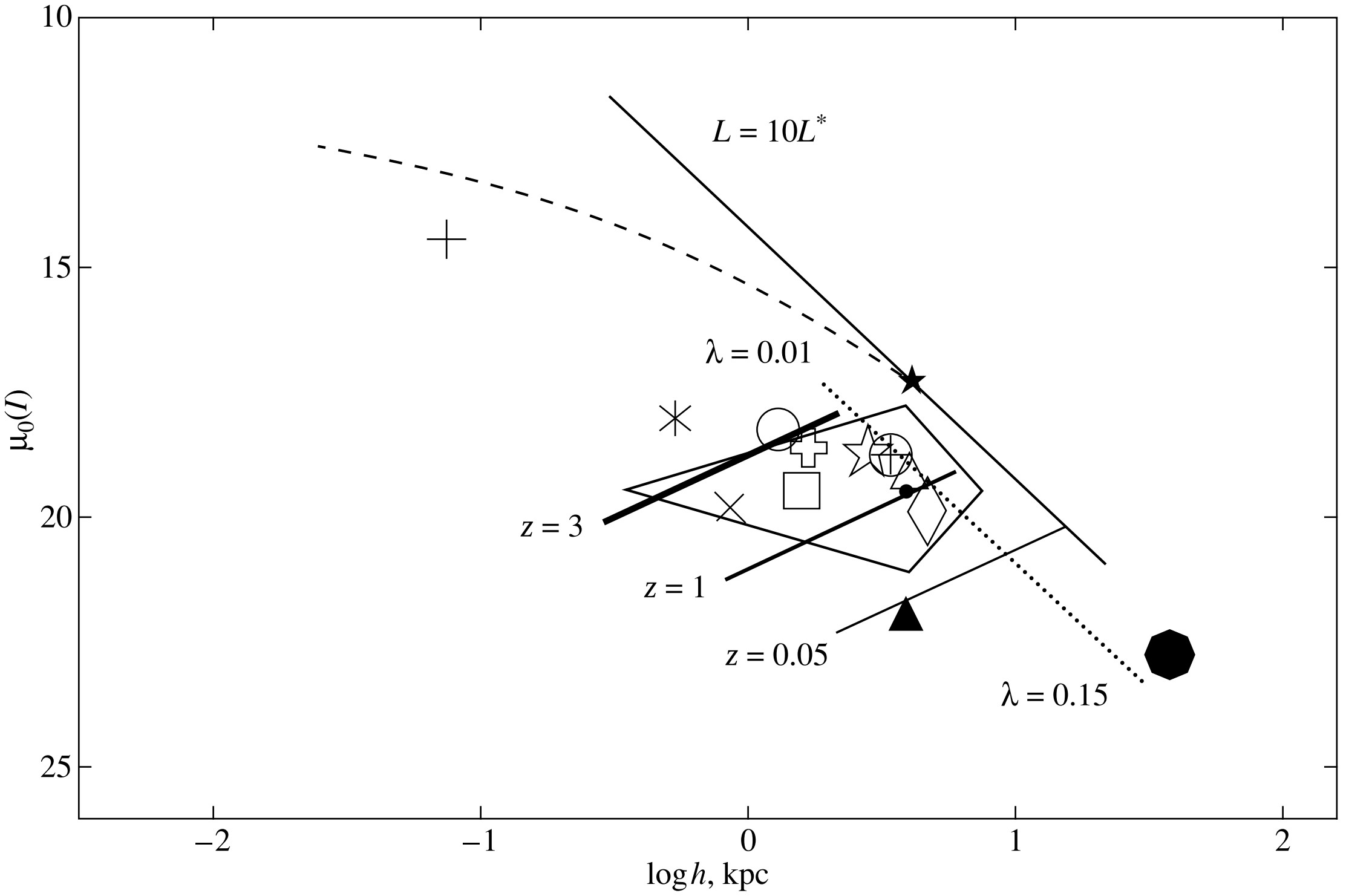

**Fig. 3.** Stellar disks of galaxies of various types on the $\mu_0(I)$–$h$ plane. The line segments of different thickness show disks loci implied by MMW models (Mo *et al.* 1998) with different formation redshifts (0.05, 1, and 3). Different symbols show the mean parameters of galaxies of various types (see Fig. 2 and text). The curves and the quadrangle have the same meaning as in Figs. 1 and 2.

The formation epoch of the Milky Way disk is by no means a very well defined quantity, because the process spanned a long time interval comparable to the Hubble time (see, e.g., Chiappini *et al.* 1997). Therefore, the epoch implied by $z_f$ should be considered only a characteristic formation age. Redshift $z_f \approx 1$ approximately corresponds to the maximum star formation rate in the Universe (Baugh *et al.* 1998).

### *4.2. Normal Galaxies*

Different symbols in Fig. 3 show the mean parameters of the galaxy samples described in Section 3 (errors are not shown in order not to encumber the figure). Stellar disks can be seen to concentrate conspicuously to the $SB_0 \propto h^{-1}$ relation. Furthermore, the mean parameters of stellar disks of galaxies of various types are mainly concentrated on the $\mu_0$–$h$ plane inside or in the vicinity of the rectangular box corresponding to normal galaxies (except compact nuclear disks and GLSB galaxies).

In Fig. 3, straight-line segments of various thickness show the parameters of stellar galaxy disks in the model of Mo *et al.* (1998) for three formation redshifts $z_f$ = 0.05, 1, and 3. We adopted $\lambda = 0.05$ and $m_d = j_d = 0.05$. The adopted spin parameter approximately corresponds to the peak of $\lambda$ distribution obtained in numerical simulations (e.g., Warren *et al.* 1992) and is close to the critical value that separates stable disks from unstable ones (Mo *et al.* 1998; van den Bosch 1998). Adopting $m_d = 0.05$ yields reasonable dark halo mass estimates in galaxies (see Section 4.1). We assumed the $j_d/m_d$ ratio to be equal to unity, which is a standard, albeit not entirely substantiated, assumption in galaxy formation models (MMW). The disk rotation velocity varies along each segment from $V_c = 50$ km s$^{-1}$ (left end) to $V_c = 350$ km s$^{-1}$ (right end).

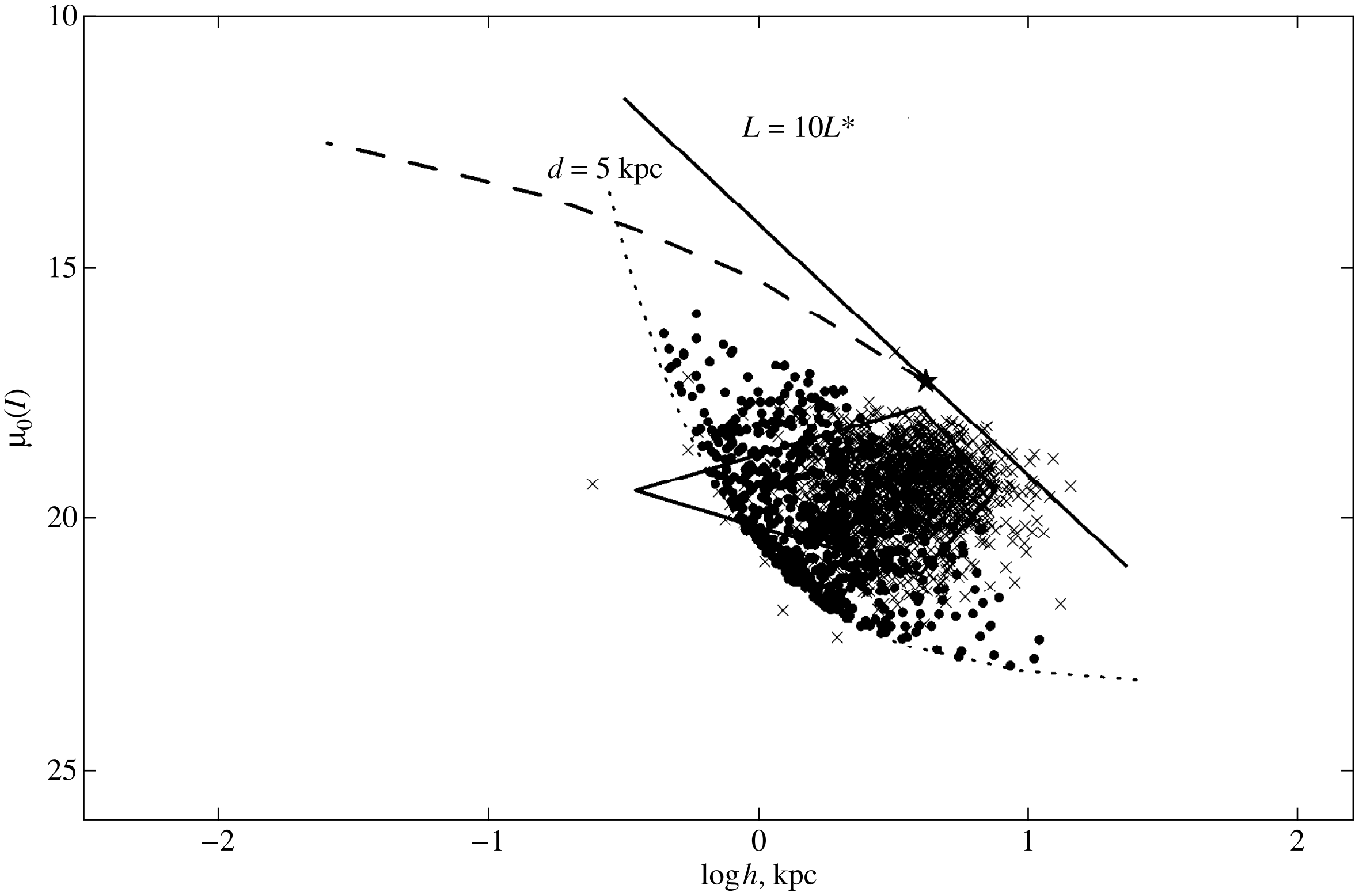

**Fig. 4.** Stellar disks of galaxies of various types on the $\mu_0(I)$–$h$ plane. Crosses show the data for galaxies from Byun's (1992) thesis; dark dots, the parameters of simulated galaxies based on the MMW model (Mo *et al.* 1998). The curves and the rectangle have the same meaning as in Figs. 1 and 2.

As is evident from Fig. 3, the MMW model explains satisfactorily both the position and scatter of observed galaxy disks on the $\mu_0$–$h$ plane. The data on normal spiral galaxies similar to the Milky Way are consistent with $z_f \sim 1$. The disks of earlier-type galaxies (*c*) have larger formation times ($z_f > 1$). The parameters of low-surface-brightness galaxies are consistent with $z_f < 1$.These estimates seem to be quite plausible if $z_f$ is considered to be the redshift of the last global star formation burst in the disk.

The dotted line in Fig. 3 shows how the parameters of the disks with $m_d = j_d = 0.05$; $V_c = 200$ km s$^{-1}$, and $z_f = 0$ vary as spin parameter $\lambda$ changes from 0.01 to 0.15. It is evident from Fig. 3 that the MMW model implies that the disks of low surface brightness galaxies are relatively "young" (i.e., unevolved) and have larger $\lambda$. This conclusion is consistent with modern theories of the objects of the type considered (e.g., Dalcanton *et al.* 1997; Jimenez *et al.* 1998).

As is evident from formulas (1)–(2), the galaxy disks must approximately obey relations $\mu_0 \propto -2.5\log V_c$ and $\log h \propto \log V_c$ (with fixed $\lambda$, $m_d$, $j_d$, and $z_f$). The data for spiral galaxies from Byun (1992) yield similar empirical relations: $\mu_0 \propto -2.72\log V_c$ ($\rho = -0.58$) and $\log h \propto 0.70 \log V_c$ ($\rho = +0.59$). It was shown in Section 2 that the MMW model implies a Tully–Fisher relation in the form $L_d \propto V_c^3$. Spiral galaxies from the thesis by Byun (1992) follow relation $L_I \propto V_c^{2.72 \pm 0.04}$ ($\rho = 0.90$).

In order to compare graphically the model with observational data, we generated a catalog of 1200 "fictitious" galaxies based on formulas (1)–(2). In our computations, we assumed that the spin parameter $\lambda$ has a log-normal distribution with $\langle\lambda\rangle = 0.05$ and $\sigma_\lambda = 0.7$ [obtained by analytically fitting the results of numerical simulations by Warren *et al.* (1992)], $V_c$ follows a generalized Schechter distribution function with $n = 3$; $\beta = -1.0$, and $V_* = 250$ km s$^{-1}$ (see Gonzalez *et al.* 1999). We further assumed that $z_f$ follows a Gaussian distribution function with $\langle z_f\rangle = 1$, $\sigma_{z_f} = 0.3$, and $m_d = j_d = 1$. In Fig. 4, we compare the parameter distributions of 705 simulated galaxies (we left only the objects with $d \geq 5$ kpc) with the corresponding distributions for the galaxies from Byun (1992). As is evident from the figure, both distributions occupy approximately the same domain on the $\mu_0$–$h$ plane. Note only a somewhat larger fractions of simulated galaxies with high and low central surface brightness $\mu_0$ compared to the corresponding fractions for real galaxies. This effect can be easily interpreted in terms of observational selection—the sample of Byun (1992) contains few low-surface brightness galaxies (they are searched for using special observational techniques) and no galaxies of early morphological subtypes. The mean parameters of the simulated galaxy sample shown in Fig. 4 ($\mu_0 = 19.9 \pm 1.3$ and $\langle h \rangle = 2.0 \pm 1.3$ kpc) agree, within the error limits, with those of the real galaxy sample from Byun *et al.* (1992). Note also that simulated galaxies follow the Freeman law (Freeman 1970) very closely.

The properties of compact stellar disks in early-type galaxies can be reproduced with small spin parameter $\lambda \sim 0.01$ and large formation times $z_f \geq 3$. However, in the galaxies with the bulge-dominated mass contribution, the disk formation mechanism is likely to differ from the standard MMW scenario. The formation of disks in such galaxies can be due, e.g., to external accretion or merging, mass loss by bulge stars, and minibar disruption, etc. (Scorza and van den Bosch, 1998; van den Bosch, 1998).

### *4.3. Distant Galaxies*

We have shown above that the global disk parameters of local spiral galaxies of various types are consistent with the predictions of the MMW model. The disks of distant galaxies that have formed long ago must be more compact and have higher surface brightnesses (MMW). Modern observational data on the photometric structure of extremely distant galaxies are, on the whole, consistent with this hypothesis.

The disks of three galaxies with $z > 5$ [HDF 4–473.0 at $z = 5.60$ according to Weymann *et al.* (1998) and the double galaxy HDF 3–951.1/2 at $z = 5.34$ according to Spinrad *et al.* (1998)] have a mean exponential scale length of $\langle h \rangle = 0.8 \pm 0.3$ kpc. The MMW model with $\lambda = 0.05$, $m_d = j_d = 0.05$, $V_c = 200$ km s$^{-1}$, and $z_f = 5$ implies $h \sim 0.6$ kpc.

The mean scale length for six galaxies located at $z \approx 3$ (Steidel *et al.* 1996; Giavalisco *et al.* 1996) is equal to $1.5 \pm 0.8$ kpc, which is comparable to the estimate implied by MMW (~1.2 kpc). The galaxies located at $z \approx 3$ have very high central surface brightnesses (in the rest frame) of $\mu_0(B) \sim 17.7$–$18.7$ (Steidel *et al.* 1996), which are about $\Delta\mu_0 = 2^m$ higher than those implied by the MMW model. This discrepancy might be due to the fact that, in the studies of distant objects, the galaxies with relatively high surface brightnesses (e.g., as a result of the ongoing star formation) are identified predominantly.

We can conclude that the MMW model is consistent with observational data on the structure of galaxies at $z \geq 3$. According to Mao *et al.* (1998), the model is also consistent with the data on the evolution of the exponential scale length $h$ at $z \leq 1 : h \propto (1 + z)^{-1}$.

## 5. CONCLUSION

We have analyzed the parameters of stellar disks in galaxies of various types and their distribution on the $\mu_0$–$h$ plane. Simple scale considerations in the CDM scenario-based disk formation model (non-self-gravitating disks embedded in isothermal dark halos—Mo *et al.* 1998) proved to satisfactorily describe both qualitatively and quantitatively the disk parameters of real galaxies and their scatter on the $\mu_0$–$h$ diagram.

In recent times, this model has often served as the basis to study various aspects of the disk galaxy formation and evolution (see, e.g., Jimenez *et al.* 1998; Boissier and Prantzos 1999; van den Bosch 1999). The fact that the estimates given by this model are consistent with observational data is really remarkable, because the approach used by Mo *et al.* (1998) ignores such important processes affecting galaxy parameters as star formation and feedback, external accretion and mergers.[2] It spite of its simplicity, the model of Mo *et al.* (1998) must reflect (statistically, i.e., in the average) some of the important factors affecting the integrated properties of the spiral galaxies in the Milky Way vicinity (see Section 2).

A serious problem of the model of Mo *et al.* (1998) (and of the CDM scenario in general) is that it overestimates the dark matter contribution inside galaxies. The mass distributions of MMW models are dark halos dominated even in the inner parts of the galaxies (Springel and White 1999), which is inconsistent with observational data for normal spirals (e.g., for the Milky Way, see Section 4.1) and agrees only with observations of low-surface-brightness galaxies. Because of the too high dark mass fraction, the CDM scenario fails to explain the zero point of the Tully–Fisher relation (see a discussion in Steinmetz and Navarro 1999). These problems must eventually result in a modification of the model of Mo *et al.* (1998).

## ACKNOWLEDGMENTS

I am grateful to N.Ya. Sotnikova (St. Petersburg State University) for her valuable comments. The work was supported by the Integration Program (project no. 578), the Competitive Center for Fundamental Science of the Ministry of General and Professional Education, and the Russian Foundation for Basic Research (project no. 98-02-18178).

[2] The effect of interactions and mergers on the structure of galaxy disks becomes considerably weaker if a later disk formation epoch of $z_f \leq 1$ is assumed (Weil *et al.* 1998)

*Translated by A. Dambis*